\documentclass[twocolumn,tighten]{aastex62}
\usepackage{color}
\usepackage{multirow}

\shortauthors{Sano, Matsumura, Yamane et al.}

\received{April 9, 2019}
\revised{May 31, 2019}
\accepted{June 17, 2019}

\begin{document}
\title{DISCOVERY OF SHOCKED MOLECULAR CLOUDS ASSOCIATED WITH THE SHELL-TYPE\\SUPERNOVA REMNANT RX~J0046.5$-$7308 IN THE SMALL MAGELLANIC CLOUD}

\author[0000-0003-2062-5692]{H. Sano}
\affiliation{Institute for Advanced Research, Nagoya University, Furo-cho, Chikusa-ku, Nagoya 464-8601, Japan; sano@a.phys.nagoya-u.ac.jp}
\affiliation{Department of Physics, Nagoya University, Furo-cho, Chikusa-ku, Nagoya 464-8601, Japan}

\author{H. Matsumura}
\affiliation{Kavli Institute for the Physics and Mathematics of the Universe (WPI), The University of Tokyo Institutes for Advanced Study, The University of Tokyo, 5-1-5 Kashiwanoha, Kashiwa, Chiba 277-8583, Japan; hideaki.matsumura@ipmu.jp}

\author[0000-0001-8296-7482]{Y. Yamane}
\affiliation{Department of Physics, Nagoya University, Furo-cho, Chikusa-ku, Nagoya 464-8601, Japan}

\author[0000-0001-5612-5185]{P. Maggi}
\affiliation{Universit\'e de Strasbourg, CNRS, Observatoire astronomique de Strasbourg, UMR 7550, F-67000 Strasbourg, France}

\author{K. Fujii}
\affiliation{Department of Astronomy, School of Science, The University of Tokyo, 7-3-1 Hongo, Bunkyo-ku, Tokyo 133-0033, Japan}

\author[0000-0002-2794-4840]{K. Tsuge}
\affiliation{Department of Physics, Nagoya University, Furo-cho, Chikusa-ku, Nagoya 464-8601, Japan}

\author[0000-0002-2062-1600]{K. Tokuda}
\affiliation{Department of Physical Science, Graduate School of Science, Osaka Prefecture University, 1-1 Gakuen-cho, Naka-ku, Sakai 599-8531, Japan}
\affiliation{National Astronomical Observatory of Japan, Mitaka, Tokyo 181-8588, Japan}

\author[0000-0001-5609-7372]{R. Z. E. Alsaberi}
\affiliation{Western Sydney University, Locked Bag 1797, Penrith South DC, NSW 1797, Australia}

\author[0000-0002-4990-9288]{M. D. Filipovi{\'c}}
\affiliation{Western Sydney University, Locked Bag 1797, Penrith South DC, NSW 1797, Australia}

\author[0000-0003-2762-8378]{N. Maxted}
\affiliation{School of {Science}, University of New South Wales, Australian Defence Force Academy, Canberra, ACT 2600, Australia}

\author[0000-0002-9516-1581]{G. Rowell}
\affiliation{School of Physical Sciences, The University of Adelaide, North Terrace, Adelaide, SA 5005, Australia}

\author[0000-0003-1518-2188]{H. Uchida}
\affiliation{Department of Physics, Graduate School of Science, Kyoto University, Kitashirakawa Oiwake-cho, Sakyo-ku, Kyoto 606-8502, Japan}

\author[0000-0002-4383-0368]{T. Tanaka}
\affiliation{Department of Physics, Graduate School of Science, Kyoto University, Kitashirakawa Oiwake-cho, Sakyo-ku, Kyoto 606-8502, Japan}

\author{K. Muraoka}
\affiliation{Department of Physical Science, Graduate School of Science, Osaka Prefecture University, 1-1 Gakuen-cho, Naka-ku, Sakai 599-8531, Japan}

\author[0000-0002-4124-797X]{T. Takekoshi}
\affiliation{Institute of Astronomy, The University of Tokyo, 2-21-1 Osawa, Mitaka, Tokyo 181-0015, Japan}
\affiliation{Graduate School of Informatics and Engineering, The University of Electro-Communications, Chofu, Tokyo 182-8585, Japan}

\author[0000-0001-7826-3837]{T. Onishi}
\affiliation{Department of Physical Science, Graduate School of Science, Osaka Prefecture University, 1-1 Gakuen-cho, Naka-ku, Sakai 599-8531, Japan}

\author[0000-0001-7813-0380]{A. Kawamura}
\affiliation{National Astronomical Observatory of Japan, Mitaka, Tokyo 181-8588, Japan}

\author[0000-0001-9778-6692]{T. Minamidani}
\affiliation{Nobeyama Radio Observatory, National Astronomical Observatory of Japan (NAOJ), National Institutes of Natural Sciences (NINS), 462-2, Nobeyama, Minamimaki, Minamisaku, Nagano 384-1305, Japan}
\affiliation{Department of Astronomical Science, School of Physical Science, SOKENDAI (The Graduate University for Advanced Studies), 2-21-1, Osawa, Mitaka, Tokyo 181-8588, Japan}

\author{N. Mizuno}
\affiliation{National Astronomical Observatory of Japan, Mitaka, Tokyo 181-8588, Japan}

\author{H. Yamamoto}
\affiliation{Department of Physics, Nagoya University, Furo-cho, Chikusa-ku, Nagoya 464-8601, Japan}

\author[0000-0002-1411-5410]{K. Tachihara}
\affiliation{Department of Physics, Nagoya University, Furo-cho, Chikusa-ku, Nagoya 464-8601, Japan}

\author{T. Inoue}
\affiliation{Department of Physics, Nagoya University, Furo-cho, Chikusa-ku, Nagoya 464-8601, Japan}

\author[0000-0003-4366-6518]{S. Inutsuka}
\affiliation{Department of Physics, Nagoya University, Furo-cho, Chikusa-ku, Nagoya 464-8601, Japan}

\author{F. Voisin}
\affiliation{School of Physical Sciences, The University of Adelaide, North Terrace, Adelaide, SA 5005, Australia}

\author[0000-0002-9931-5162]{N. F. H. Tothill}
\affiliation{Western Sydney University, Locked Bag 1797, Penrith South DC, NSW 1797, Australia}

\author[0000-0001-5302-1866]{M. Sasaki}
\affiliation{Dr. Karl Remeis-Sternwarte, Erlangen Centre for Astroparticle Physics, Friedrich-Alexander-Universit$\ddot{a}$t Erlangen-N$\ddot{u}$rnberg, Sternwartstra$\beta$e 7, D-96049 Bamberg, Germany}

\author[0000-0003-2730-957X]{N. M. McClure-Griffiths}
\affiliation{Research School of Astronomy \& Astrophysics, The Australian National University, Canberra, ACT 2611, Australia}

\author{Y. Fukui}
\affiliation{Institute for Advanced Research, Nagoya University, Furo-cho, Chikusa-ku, Nagoya 464-8601, Japan; sano@a.phys.nagoya-u.ac.jp}
\affiliation{Department of Physics, Nagoya University, Furo-cho, Chikusa-ku, Nagoya 464-8601, Japan}

\begin{abstract}
RX~J0046.5$-$7308 is a shell-type supernova remnant (SNR) in the Small Magellanic Cloud (SMC). We carried out new $^{12}$CO($J$ = 1--0, 3--2) observations toward the SNR using Mopra and the Atacama Submillimeter Telescope Experiment. We found eight molecular clouds (A--H) along the X-ray shell of the SNR. The typical cloud size and mass are $\sim$10--15~pc and $\sim$1000--3000~$M_\sun$, respectively. The X-ray shell is slightly deformed and has the brightest peak in the southwestern shell where two molecular clouds A and B are located. The four molecular clouds A, B, F, and G have {high intensity ratios} of $^{12}$CO($J$ = 3--2) / $^{12}$CO($J$ = 1--0) $> 1.2$, which are not attributable to any identified internal infrared sources or high-mass stars. The H{\sc i} cavity and its expanding motion are found toward the SNR, which are likely created by strong stellar winds from a massive progenitor. We suggest that the molecular clouds A--D, F, and G, and H{\sc i} clouds within the wind-blown cavity at $V_\mathrm{LSR} = 117.1$--122.5 km s$^{-1}$ are associated with the SNR. The X-ray spectroscopy reveals the dynamical age of { $26000^{+1000}_{-2000}$ yr} and the progenitor mass of {$\gtrsim 30~M_{\sun}$}, which is also consistent with the proposed scenario. We determine physical conditions of the giant molecular cloud LIRS~36A using the large velocity gradient analysis with archival datasets of the Atacama Large Millimeter/submillimeter Array; the kinematic temperature is $72^{+50}_{-37}$~K and the number density of molecular hydrogen is $1500^{+600}_{-300}$~cm$^{-3}$. The next generation of $\gamma$-ray observations will allow us to study the pion-decay $\gamma$-rays from the {molecular clouds} in the SMC SNR.
\end{abstract}

\section{Introduction}
In our Galaxy, molecular clouds associated with supernova remnants (SNRs) play an essential role in understanding not only the shock heating/compression of the interstellar medium (ISM), but also the origins of thermal/nonthermal X-rays, $\gamma$-rays, and cosmic rays. The shock--cloud interaction excites turbulence that enhances the magnetic field up to $\sim$1~mG \citep[e.g.,][]{2007Natur.449..576U,2009ApJ...695..825I,2012ApJ...744...71I}, which can be observed as shocked gas clumps with limb brightening in the synchrotron X-rays \citep[e.g.,][]{2010ApJ...724...59S,2013ApJ...778...59S,2018PASJ...70...77O}. For the ionized plasma in the Galactic SNR RCW~86, \cite{2017JHEAp..15....1S} found a positive correlation between the thermal X-ray flux and the gas density around the shocked region, indicating that shock ionization occurred. The interstellar protons also act as a target for cosmic-ray protons producing GeV/TeV $\gamma$-rays via neutral pion decay. The good spatial correspondence between the interstellar protons and $\gamma$-rays provides evidence for cosmic-ray acceleration in the Galactic SNRs \citep[e.g.,][]{2003PASJ...55L..61F,2008A&A...481..401A,2012ApJ...746...82F,2017ApJ...850...71F,2013ApJ...768..179Y,2013ASSP...34..249F}.

The Magellanic Clouds---consisting of the Large Magellanic Cloud (LMC) and Small Magellanic Cloud (SMC)---provide us with unique laboratories for studying the shock interaction because of their well-known distance ($50 \pm 1.3$ kpc for the LMC, \citeauthor{2013Natur.495...76P} \citeyear{2013Natur.495...76P}; $\sim60$ kpc for the SMC, \citeauthor{2005MNRAS.357..304H} \citeyear{2005MNRAS.357..304H}) and low ISM metallicity ($\sim$0.3--0.5~$Z_\sun$ for the LMC, \citeauthor{1997macl.book.....W} \citeyear{1997macl.book.....W}; $\sim${0.05--0.2}~$Z_\sun$ for the SMC, \citeauthor{1992ApJ...384..508R} \citeyear{1992ApJ...384..508R}; \citeauthor{1999AA...348..728R} \citeyear{1999AA...348..728R}). The smaller contamination along the line-of-sight is also {{advantageous}} to identify molecular clouds associated with the SNRs. For LMC SNRs N23, N49, and N132D, \cite{1997ApJ...480..607B} carried out pioneering CO studies by using the Swedish-ESO Submillimetre Telescope. Recent CO observations using the Atacama Submillimeter Telescope Experiment (ASTE), Mopra, and Atacama Large Millimeter/submillimeter Array (ALMA) revealed clumpy molecular clouds associated with the X-ray bright LMC SNRs with an angular resolution of $2''$--$45''$, corresponding to the spatial resolution of 0.5--11 pc \citep{2015ASPC..499..257S,2017AIPC.1792d0038S,2017ApJ...843...61S,2018ApJ...867....7S,2019ApJ...873...40S,2018ApJ...863...55Y}. Most recently, \cite{arXiv:1903.03226} {{found}} an H{\sc i} cavity interacting with the SMC SNR DEM~S5. The H{\sc i} data was obtained using the Australian Square Kilometer Array Pathfinder (ASKAP) with an angular resolution of $\sim$$30''$, corresponding to the spatial resolution of $\sim$9 pc {at} the SMC distance. We are, therefore, entering a new age in studying the Magellanic SNRs that yield a spatial resolution comparable to what had been possible only for Galactic SNRs. However, there are no CO observations toward the SMC SNRs.

RX~J0046.5$-$7308 (also known as SNR~B0044$-$73.4{,} HFPK~414{, or DEM~S32}) is an X-ray SNR located in the southwestern part of the SMC \citep[e.g.,][]{2000AAS..142...41H,2004AA...421.1031V,2008A&A...485...63F,2012A&A...545A.128H,2015ApJ...803..106R}. The description of the source first appeared in the X-ray survey paper of the SMC \citep{1992ApJS...78..391W}. Subsequent optical, radio continuum, and X-ray observations confirmed that RX~J0046.5$-$7308 is a shell-type SNR in the vicinity of the H{\sc ii} region N19 \citep[e.g.,][]{1994AA...286..231R,2001AJ....122..849D,2004AA...421.1031V,2005MNRAS.364..217F,2007MNRAS.376.1793P}. The size of X-ray shell is about 40.7 pc $\times$ 46.5 pc \citep{2008A&A...485...63F}, which spatially coincides with the radio continuum shell with a spectral index of $-0.6$ \citep[e.g.,][]{2001AJ....122..849D,2005MNRAS.364..217F}. Recently, a spectral index of $-0.38 \pm 0.05$ {{suggesting a flatter spectrum}} was derived, indicating the thermal origin of the radio continuum and an {evolved} SNR (Maggi et al., in preparation). In fact, {in previous $XMM$-$Newton$ studies} the X-ray spectra were well described by a simple nonequilibrium ionization (NEI) plasma model without synchrotron X-rays \citep[e.g.,][]{2004AA...421.1031V}. Assuming the Sedov model, the ionization age was estimated as $\sim$15,000~yr. Although the progenitor type could not be determined , the rich {star-forming} environment (N19) might suggest that RX~J0046.5$-$7308 is a core-collapse (CC) SNR. This means that the SNR has a potential to be associated with dense molecular clouds. \cite{1993AA...271....1R,1993A&A...271....9R} revealed a giant molecular cloud (GMC) located in the north of the SNR, corresponding to the H{\sc ii} region DEM~S23 (also known as N12A or NGC~261). The fundamental physical properties have been derived: e.g., a virial mass of $\sim$17000~$M_\sun$ and a radius of $\sim$18.6~pc \citep{1993A&A...271....9R,1994AA...292..371L,2007AA...471..561N}, but the physical relation between the GMC and SNR has not been discussed. {The issue is further complicated by the large depth along the line of sight of the SMC \citep{Scowcroft_2016}, making GMC/SNR association less likely based on projected location only.}

In {this} paper, we report the first detection of molecular clouds and atomic gas associated with the SMC SNR~RX~J0046.5$-$7308 using the ASTE, Mopra, and ASKAP. We also present the physical properties of the GMC in DEM~S23 and its relation with the SNR using archival CO data taken with ALMA. Section \ref{sec:obs} describes observations and data reductions of CO, H{\sc i}, and X-rays. Section \ref{largescale} gives large-scale maps of X-rays and CO; Sections \ref{physicalproperties}, \ref{LIRS36AB}--\ref{ratio} describe physical properties of molecular clouds; Section \ref{hi_cloud} presents H{\sc i} distribution; and Section \ref{sec:xspec} gives a X-ray spectral analysis. Discussion and conclusions are provided in Sections \ref{sec:discussion} and \ref{sec:conclusions}.

\begin{figure*}[]
\begin{center}
\includegraphics[width=\linewidth]{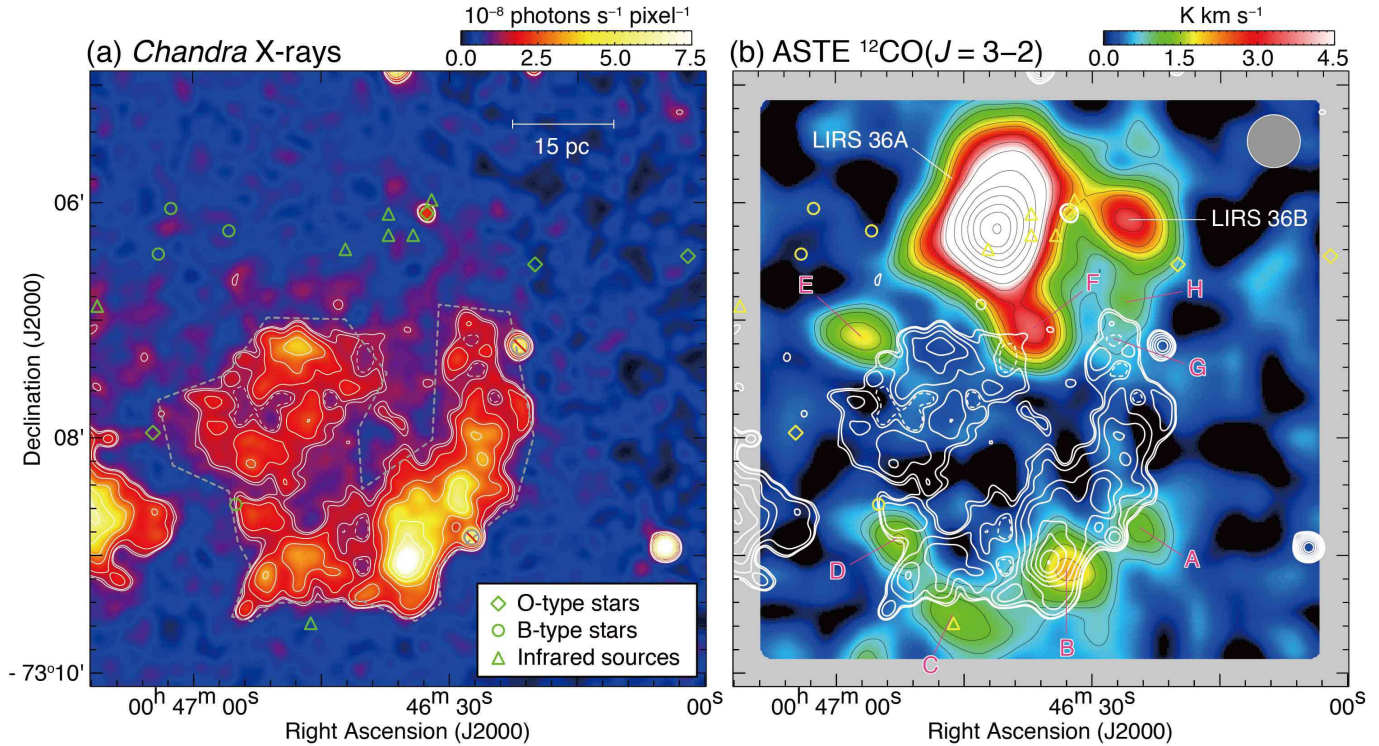}
\caption{(a) {\it{Chandra}} X-ray image of RX~J0046.5$-$7308 in the energy band of 0.3--2.1 keV. The contour levels are 1.30, 1.43, 1.80, 2.43, 3.30, 4.43, 5.80, 7.43, and $9.3 \times 10^{-8}$ photons s$^{-1}$ pixel$^{-1}$. The region enclosed by dashed lines is used for the X-ray spectral analysis. The scale bar is also shown in top right corner of Figure \ref{fig1}(a). (b) Integrated intensity map of ASTE $^{12}$CO($J$ = 3--2) overlaid with the {\it{Chandra}} X-ray intensity as shown in Figure \ref{fig1}(a) (white contours). The integration velocity range is from $V_\mathrm{LSR}$ = 117.1 to 130.1 km s$^{-1}$. The black contours indicate $^{12}$CO($J$ = 3--2) integrated intensity, whose contour levels are 0.7, 0.9, 1.1, 1.5, 1.9, 2.3, 3.1, 3.9, 4.7, 6.3, 7.9, 9.5, 11.1, 12.7, and 14.3 K km s$^{-1}$. CO peaks A--H and GMCs LIRS~36A--B discussed in Section 3 are indicated in the figure. The beam size of the ASTE is also shown in top right corner of Figure \ref{fig1}(a). The diamond, circle, and triangle symbols represent the positions of O-type stars, B-type stars, and infrared sources, respectively {\citep{2003yCat.2246....0C,2003A&A...401..873W,2018yCat.1345....0G}}.}
\label{fig1}
\end{center}
\end{figure*}%

\section{Observations and Data Reductions}\label{sec:obs}
\subsection{CO}
{To investigate shocked molecular clouds associated with the SNR RX~J0046.5$-$7308, we observed both the $^{12}$CO($J$ = 3--2) and $^{12}$CO($J$ = 1--0) line emission. The intensity ratio between the $^{12}$CO($J$ = 3--2) and $^{12}$CO($J$ = 1--0) transitions is a good indicator {{for}} shock-heated molecular clouds (for more information, see Section \ref{section4.2}).}

Observations of $^{12}$CO($J$ = 3--2) line emission at 345.795990~GHz were carried out during 2014 August by using the ASTE 10 m radio telescope \citep{2004SPIE.5489..763E}, which was operated by the National Astronomical Observatory of Japan (NAOJ). We observed $5' \times 5'$ rectangular region centered at ($\alpha_\mathrm{J2000}$, $\delta_\mathrm{J2000}$) $=$ ($00^\mathrm{h}46^\mathrm{m}36$\farcs$02$, $-73\degr07\arcmin30\farcs1$) using the on-the-fly (OTF) mapping mode with Nyquist sampling. The front end was a side band separating the superconductor-insulator-superconductor mixer receiver {``CATS345''} \citep{2008stt..conf..281I}. We utilized a digital FX spectrometer ``MAC'' \citep{2000SPIE.4015...86S} as the back end. The bandwidth of the spectrometer is 128~MHz with 1024~channels, corresponding to the velocity coverage of $\sim$111 km s$^{-1}$ and the velocity resolution of $\sim$0.11 km s$^{-1}$. The typical system temperature was $\sim$200--300~K, including the atmosphere in the single-side band (SSB). To derive the main beam efficiency, we observed N12A {[($\alpha_\mathrm{J2000}$, $\delta_\mathrm{J2000}$) $=$ ($00^\mathrm{h}46^\mathrm{m}41$\farcs$4$, $-73\degr06\arcmin05''$)]} \citep{2007AA...471..561N} and obtained {{a}} main beam efficiency of $\sim$0.71. The pointing accuracy was checked every half an hour to achieve an offset within $2''$. After smoothing with a two-dimensional Gaussian kernel, we obtained {{a}} data cube with the beam size of $\sim$$27''$ ($\sim$8 pc at the {distance to the} SMC). The typical noise fluctuation is $\sim$0.038 K at the velocity resolution of 0.4 km s$^{-1}$.

\begin{figure*}[]
\begin{center}
\includegraphics[width=\linewidth]{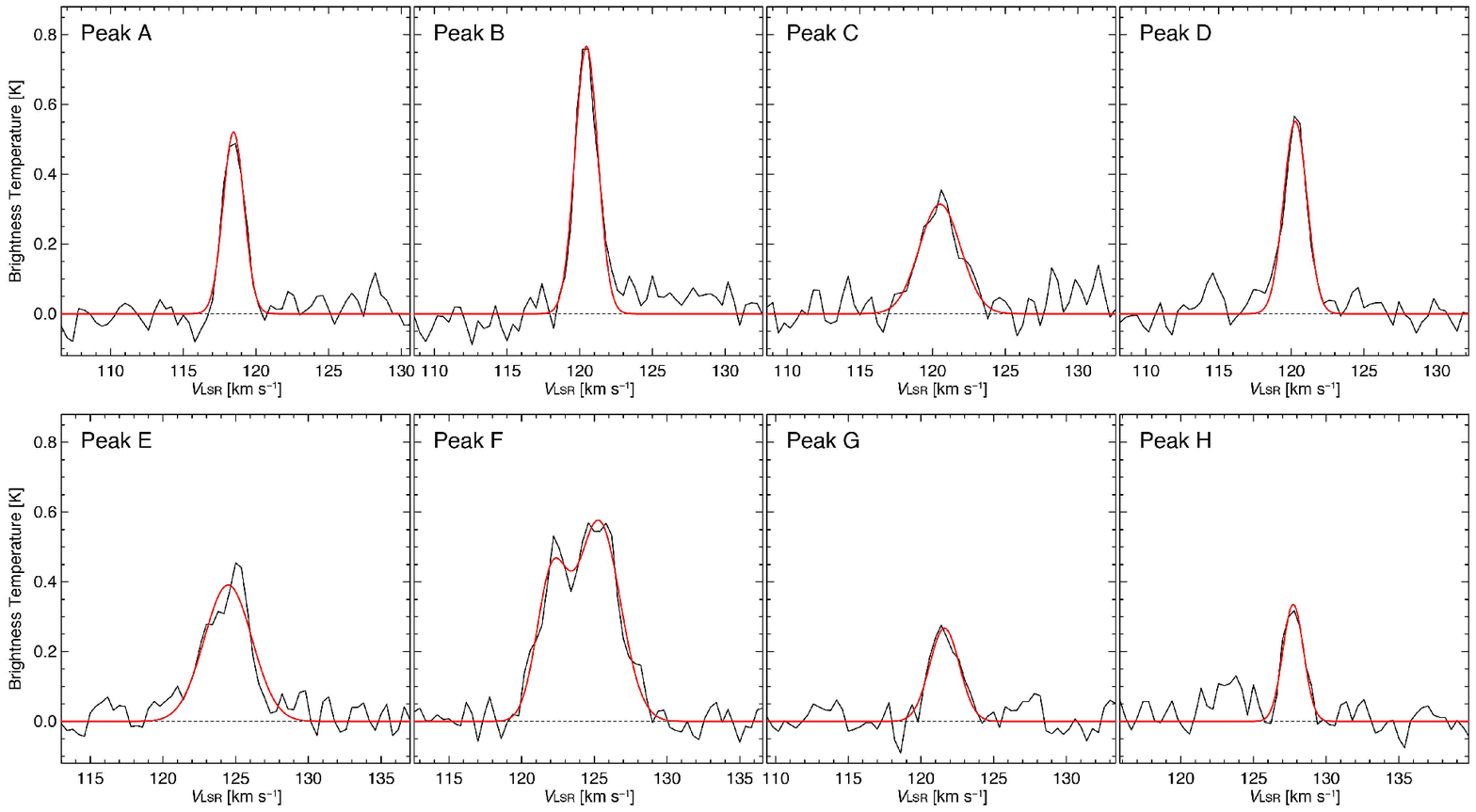}
\caption{$^{12}$CO($J$ = 3--2) profiles toward CO peaks A--H (black solid lines). The red lines indicate least-squares fitting results using a single Gaussian function (peaks A--E, G, and F) or a double Gaussian function (peak F).}
\label{fig2}
\end{center}
\end{figure*}%

Observations of $^{12}$CO($J$ = 1--0) line emission at 115.271202~GHz were executed from 2014 July to September using the Mopra 22~m radio telescope of the Commonwealth Scientific and Industrial Research Organization (CSIRO). We used the OTF mapping mode with Nyquist sampling. The map size and center position are the same as that of $^{12}$CO($J$ = 3--2). The front end was an Indium Phosphide (InP) High Electron Mobility Transistor receiver (HEMT). We utilized a digital filter-bank spectrometer (MOPS) system as the back end. The spectrometer has 4096 channels with the bandwidth of 137.5~MHz, corresponding to the velocity resolution of $\sim$0.1 km s$^{-1}$ and the velocity coverage of $\sim$360 km s$^{-1}$. The typical system temperature was $\sim$700--800~K including the atmosphere in the SSB. The pointing accuracy was checked every 2 hr and was {{found}} to be within an offset of $\sim$$5''$. We also derived the main beam efficiency of $\sim$0.46 by observing Ori-KL [($\alpha_\mathrm{J2000}$, $\delta_\mathrm{J2000}$) $=$ ($05^\mathrm{h}35^\mathrm{m}38$\farcs$6$, $-05\degr22\arcmin30''$)] \citep{2005PASA...22...62L} as the absolute intensity calibrator. After two-dimensional Gaussian smoothing, we obtained  data with {{a}} beam size of $\sim$$45''$ ($\sim$13~pc at the SMC). Finally, we combined the data with archival Mopra data ``MAGMA-SMC'' \citep{2013IAUS..292..110M} using the rms weighting scheme. The final noise fluctuation of the data was $\sim$0.067~K at the velocity resolution of 0.53 km s$^{-1}$.

To derive the density and kinematic temperature of the GMC LIRS~36, we used archival CO datasets obtained with ALMA Band~3 (86--116~GHz) and Band~6 (211--275~GHz) as a Cycle~3 project \#2015.1.00196.S (PI: J. Roman-Duval). Observations of $^{12}$CO($J$ = 2--1) line emission at 230.538000~GHz and $^{13}$CO($J$ = 2--1) line emission at 220.398684~GHz were conducted in June and 2016 August using three antennas of a total power (TP) array. The OTF mapping mode with Nyquist sampling was used. The map size is about a $2' \times 2'$ rectangular region centered at the GMC LIRS~36 [($\alpha_\mathrm{J2000}$, $\delta_\mathrm{J2000}$) $=$ ($00^\mathrm{h}46^\mathrm{m}41$\farcs$50$, $-73\degr06\arcmin00''$)]. We utilized the product dataset through the pipeline processes using the Common Astronomy Software Application \citep[CASA;][]{2007ASPC..376..127M} package version 4.5.3 with the Pipeline version r36660 (Pipeline-Cycle3-R4-B). The beam size is $\sim$$29.1''$ for $^{12}$CO($J$ = 2--1) and is $\sim$$30.3''$ for $^{13}$CO($J$ = 2--1), corresponding to a spatial resolution of $\sim$9~pc at the distance of the SMC. The typical noise fluctuations of the $^{12}$CO($J$ = 2--1) and $^{13}$CO($J$ = 2--1) data are $\sim$0.011~K and $\sim$0.014~K at the velocity resolution of 0.32 km s$^{-1}$, respectively.

\begin{deluxetable*}{lccccccccccc}[]
\tablecaption{Properties of CO clouds associated with RX~J0046.5$-$7308}
\tablehead{\multicolumn{1}{c}{Name} & $\alpha_{\mathrm{J2000}}$ & $\delta_{\mathrm{J2000}}$ & $T_{\mathrm{peak}} $ & $V_{\mathrm{peak}}$ & $\Delta V$ & Size  & $M_\mathrm{CO}$ & {$R_\mathrm{CO32/CO10}$} & Comment\\
& ($^{\mathrm{h}}$ $^{\mathrm{m}}$ $^{\mathrm{s}}$) & ($^{\circ}$ $\arcmin$ $\arcsec$) & (K) & \scalebox{0.9}[1]{(km $\mathrm{s^{-1}}$)} & \scalebox{0.9}[1]{(km $\mathrm{s^{-1}}$)} & (pc) &  ($M_\sun $) & &\\
\multicolumn{1}{c}{(1)} & (2) & (3) & (4) & (5) & (6) & (7) & (8) & (9) & (10) }
\startdata
A & 00 46 24.5 & $-73$ 08 50 & 0.52 & 118.5 & 1.7 & 12.7 & \phantom{0}\phantom{0}1300 & {1.0} & -----\\
B & 00 46 33.7 & $-73$ 09 10 & 0.77 & 120.5 & 1.9 & 16.7 & \phantom{0}\phantom{0}2600 & {1.1} & -----\\
C & 00 46 45.2 & $-73$ 09 30 & 0.31 & 120.5 & 3.3 & 13.9 & \phantom{0}\phantom{0}2400 & {0.8} & -----\\
D & 00 46 52.1 & $-73$ 09 00 & 0.55 & 120.3 & 1.9 & 12.3 & \phantom{0}\phantom{0}1000 & {1.5} & -----\\
E & 00 46 56.7 & $-73$ 07 10 & 0.39 & 124.4 & 3.9 & 11.4 & \phantom{0}\phantom{0}2600 & {0.7} & -----\\
F & 00 46 36.0 & $-73$ 07 10 & 0.40/0.57 & 122.1/125.3 & 2.5/5.0 &15.7 & \phantom{0}\phantom{0}5800 & 1.0 & double peaks\\
G & 00 46 26.8 & $-73$ 07 10 & 0.27 & 121.6 & 2.4 & \phantom{0}9.3 & \phantom{0}$ > 300$ & {1.9} & ----- \\
H & 00 46 24.5 & $-73$ 06 40 & 0.34 & 127.7 & 1.7 & 13.9 & \phantom{0}\phantom{0}1700 & {0.8} &  -----\\
LIRS~36A & 00 46 40.6 & $-73$ 06 10 & 4.15 & 126.3 & 3.0 & 32.2  &\phantom{0}37000 & {1.1} & LIRS~36 main\\
LIRS~36B & 00 46 24.6 & $-73$ 06 10 & 0.82 & 122.6 & 3.4 & 17.4 &  \phantom{0}\phantom{0}3800 & {1.3} & LIRS~36 sub\\
\enddata
\tablecomments{Col. (1): CO cloud name. Cols. (2)--(6): observed properties of the CO cloud obtained by single or double Gaussian fitting with $^{12}$CO($J$ = 3--2) emission line. Cols. (2)--(3): positions of the CO peak intensity. Col. (4): maximum brightness temperature. Col. (5): central velocity. Col. (6): linewidth $\Delta V$ (FWHM). Col. (7): CO cloud size defined as $(S / \pi)^{0.5} \times  2$, where $S$ is the CO cloud surface area surrounded by contours of the $8\sigma$ level. Col. (8): CO cloud mass $M_\mathrm{CO}$ derived by using an equation of $N(\mathrm{H_2}) / W(\mathrm{CO}) = 7.5 \times 10^{20}$ (K km s$^{-1}$)$^{-1}$ cm$^{-2}$, where $N(\mathrm{H_2})$ is molecular hydrogen column density and $W(\mathrm{CO})$ is the integrated intensity of $^{12}$CO($J$ = 1--0) \citep{2017ApJ...844...98M}. {$W(\mathrm{CO})$ was derived from the integrated intensity of $^{12}$CO($J$ = 3--2) using the intensity ratio of $^{12}$CO($J$ = 3--2) / $^{12}$CO($J$ = 1--0) for each cloud.} {Col. (9): intensity ratio of $^{12}$CO($J$ = 3--2) / $^{12}$CO($J$ = 1--0) for each cloud. Col. (10)}: name of the CO cloud LIRS~36 identified in \cite{1993AA...271....1R} is also noted. }
\label{table}
\end{deluxetable*}

\subsection{H{\sc i}}
We used H{\sc i} data published by \cite{2018NatAs...2..901M} and \cite{2019MNRAS.483..392D}. The H{\sc i} data were obtained using the ASKAP \citep{2009IEEEP..97.1507D}. The angular resolution of the H{\sc i} data is 35\farcs03 $\times$ 26\farcs96 with a position angle of 89\fdg62, corresponding to the spatial resolution of $\sim$9 pc at the SMC. The typical noise fluctuations of the H{\sc i} is $\sim$0.7 K at the velocity resolution of 3.9 km s$^{-1}$.

\subsection{X-rays}
We used archival X-ray data obtained {by} using {\it{Chandra}}, for which the observation IDs (Obs IDs) are 3904 (PI: R.~Williams), 14674, 15507, and 16367 (PI: A.~Zezas), which have been published {{by previous authors}} \citep[e.g.,][]{2006ESASP.604..375W,2008ApJS..177..216G,2008MNRAS.389..806S,2016ApJ...829...30C,2016MNRAS.462.4371I,2017AN....338..220Y,2017ApJ...839..119Y,2017RAA....17...59C,2017ApJ...847...26H,2018A&A...614A..34D}. The datasets were taken with the Advanced CCD Imaging Spectrometer S-array {(ACIS-S2)} on 2003 January for Obs ID 3904 and with the ACIS-I array on 2013 March and September for Obs IDs 14674, 15507, and 16367. We utilized {{\it{Chandra}}} Interactive Analysis of Observations \citep[CIAO;][]{2006SPIE.6270E..1VF} software version 4.10 with CALDB~4.7.8 for data reduction and imaging. The datasets were reprocessed using the chandra\_repro tool. We created an energy-filtered, exposure-corrected image using the fluximage tool in the energy band 0.3--2.1~keV. The total effective exposure is $\sim 168.8$~ks. We finally smoothed the data with a Gaussian kernel of $6\arcsec$ (FWHM). For the spectral analysis, we reduced and processed the data using the HEADAS software version 6.19. {We created spectra with the four {\it Chandra} data using the specextract tools. The ACIS-I spectra were combined by using the combine\_spectra tool.} {We did not combine the ACIS-S and ACIS-I spectra since they have different responses (intrinsically and because of increased filter contamination over the 10 years separating the two sets of data).} All X-ray spectral fits are performed with XSPEC version 12.10.0e. The plasma models are calculated with ATOMDB version 3.0.9 with the solar abundance given by \cite{Wilms2000}. The errors are quoted at 1$\sigma$ confidence levels in the text, tables, and figures in the X-ray analysis.

\section{Results}\label{sec:results}
\subsection{Large-scale Distribution of the CO and X-rays}\label{largescale}
Figure \ref{fig1}(a) shows an X-ray image of RX~J0046.5$-$7308 obtained with {\it Chandra} in the energy band of 0.3--2.1~keV. The {incomplete X-ray shell with a possible blowout feature toward the northwest} is spatially resolved. The spatial extent of the SNR is about 55~pc $\times$ 45~pc, which is roughly consistent with the previous radio continuum and X-ray studies \citep[e.g.,][]{2001AJ....122..849D,2004AA...421.1031V,2008A&A...485...63F}. We find multiple local peaks of X-rays {{in}} the shell; the brightest X-ray peak appears in the southwestern rim. We also note that an X-ray point source ($\alpha_\mathrm{J2000}$, $\delta_\mathrm{J2000}$) $=$ ($00^\mathrm{h}46^\mathrm{m}32$\farcs$63$, $-73\degr06\arcmin05\farcs7$) coincides with the O3/4V-type star LIN~78 {\citep{2003yCat.2246....0C,2018yCat.1345....0G}}, which is an exciting star of DEM~S23. An X-ray source ($\alpha_\mathrm{J2000}$, $\delta_\mathrm{J2000}$) $\sim$ ($00^\mathrm{h}47^\mathrm{m}00^\mathrm{s}$, $-73\degr08\arcmin40''$) is {on the} edge of another SNR RX~J0047.5$-$7308 (also known as IKT~2 or MCSNR~J0047$-$7308).

Figure \ref{fig1}(b) shows a large-scale distribution of $^{12}$CO($J$ = 3--2) toward the SNR RX~J0046.5$-$7308. We discovered eight molecular clouds, A--H, along the X-ray shell: four of them (A--D) delineate the southern shell, while the others (E--H) are located in outer boundaries of the northern shell. Clouds C and D are possibly associated with an infrared source and a B-type star {\citep{2003A&A...401..873W,2018yCat.1345....0G}}. We also find complementary spatial distributions between the molecular clouds and X-ray peaks, especially in cloud B. The typical separation between the peaks of CO and X-rays is from a few pc to 10~pc.

\subsection{Physical properties of eight molecular clouds}\label{physicalproperties}
Figure \ref{fig2} shows line profiles of CO. All molecular clouds were significantly detected in $^{12}$CO($J$ = 3--2), which are well described by a single Gaussian model for peaks A--E, G, and F or a double Gaussian model for peak F. CO peaks C, E, F, and G show slightly larger linewidths $> 2.5$ km s$^{-1}$, but we could not find reliable evidence of shock-broadening or winglike profiles of CO, possibly due to the coarse angular resolution of the current datasets. The properties of CO clouds (position, peak intensity/velocity, linewidth, size, and mass) are summarized in Table \ref{table}.

For the mass estimation, we used the CO-derived mass, $M_\mathrm{CO}$, using the following equation:
\begin{eqnarray}
M_{\mathrm{CO}} = m_{\mathrm{H}} \mu D^2 \Omega \sum_{i} [N_i(\mathrm{H}_2)],\\
N(\mathrm{H}_2) = X_\mathrm{CO} \cdot W(\mathrm{CO}),
\label{eq2}
\end{eqnarray}
where $m_{\mathrm{H}}$ is the atomic hydrogen mass, $\mu$ is the mean molecular weight of $\sim$2.7, $D$ is the distance to the SMC {in units of cm}, $\Omega$ is the solid angle of each pixel, $N_i(\mathrm{H}_2)$ is the column density of molecular hydrogen for each pixel $i$ {in units of cm$^{-2}$}, $X_\mathrm{CO}$ is the CO-to-H$_2$ conversion factor {in units of (K km s$^{-1}$)$^{-1}$ cm$^{-2}$}, and $W(\mathrm{CO})$ is the integrated intensity of $^{12}$CO($J$ = 1--0) line emission {in units of K km s$^{-1}$}. In the present paper, we used $X_\mathrm{CO} = 7.5 \times 10^{20}$ (K km s$^{-1}$)$^{-1}$ cm$^{-2}$ \citep{2017ApJ...844...98M}. To calculate $W(\mathrm{CO})$, we used the ASTE $^{12}$CO($J$ = 3--2) data instead of the Mopra $^{12}$CO($J$ = 1--0) data, because the $^{12}$CO($J$ = 3--2) data have a higher angular resolution and {better} sensitivity than the $^{12}$CO($J$ = 1--0) data. We converted from the integrated intensity of $^{12}$CO($J$ = 3--2) into $W(\mathrm{CO})$ using the intensity ratio of $^{12}$CO($J$ = 3--2) / $^{12}$CO($J$ = 1--0) (hereafter  $R_\mathrm{CO32/CO10}$) for each cloud. The typical cloud size and CO-derived mass are $\sim$10--15~pc and $\sim$1000--3000~$M_\sun$, respectively.

\begin{figure}[]
\begin{center}
\vspace*{0.5cm}
\includegraphics[width=\linewidth]{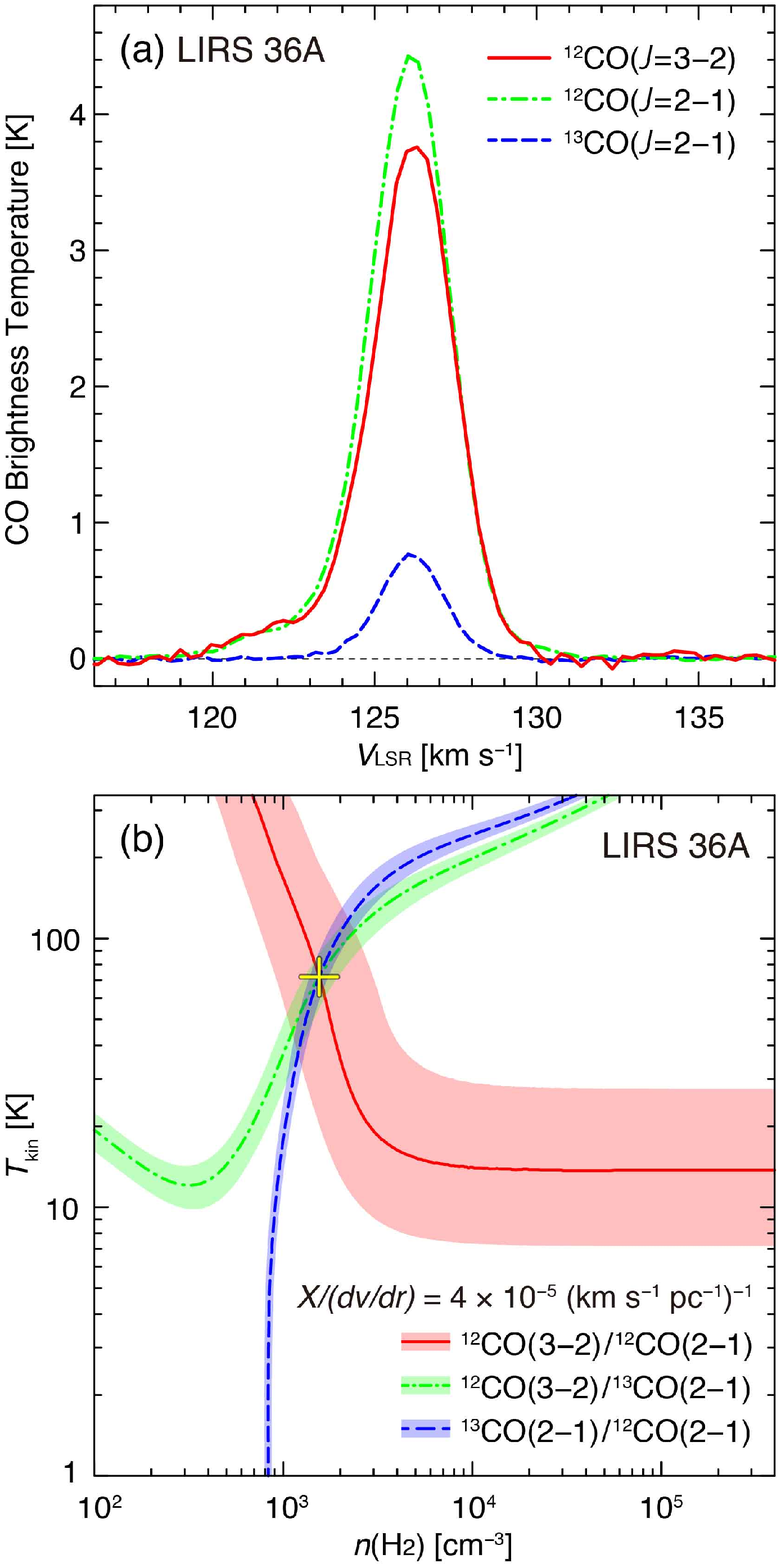}
\caption{(a) CO spectra toward LIRS~36A. The red, green, and black spectra represent the ASTE $^{12}$CO($J$ = 3--2), ALMA $^{12}$CO($J$ = 2--1), and ALMA $^{13}$CO($J$ = 2--1) emission lines, respectively. (b) LVG results on the number density of molecular hydrogen, $n$(H$_2$), and the kinematic temperature, $T_\mathrm{kin}$, plane toward LIRS~36A. The red, green, and blue solid lines represent the intensity ratios of $^{12}$CO($J$ = 3--2)/$^{12}$CO($J$ = 2--1), $^{12}$CO($J$ = 3--2)/$^{13}$CO($J$ = 2--1), and $^{13}$CO($J$ = 2--1)/$^{12}$CO($J$ = 2--1), respectively. The cross indicates the best-fit values of $n$(H$_2$) and $T_\mathrm{kin}$ (for details, see the text).}
\label{fig3}
\end{center}
\end{figure}%

\subsection{The GMCs of LIRS~36A and LIRS~36B}\label{LIRS36AB}
Two GMCs---hereafter referred to as ``LIRS~36A'' and ``LIRS~36B''---are also detected toward the north of the SNR (see Figure \ref{fig1}(b)) and were named as a single cloud, ``LIRS~36,'' by the previous CO studies \citep[e.g.,][]{1993A&A...271....9R,1996A&AS..118..263R,2007AA...471..561N}. The GMCs may not be strongly related with the SNR owing to their large separation from the northern shell boundary but are possibly influenced by ultraviolet (UV) radiation from the massive star in the H{\sc ii} region DEM~S23 because of their low-metal environment \citep[e.g.,][]{2003A&A...406..817I}. To reveal detailed physical conditions of the GMC LIRS~36A, we performed the large velocity gradient (LVG) analysis \citep[e.g., ][]{1974ApJ...189..441G,1974ApJ...187L..67S} using the ALMA $^{12}$CO($J$ = 2--1) and $^{13}$CO($J$ = 2--1) data and the ASTE $^{12}$CO($J$ = 3--2) data. The $^{13}$CO line data {are} suitable for deriving the velocity gradient because they trace the densest part of the cloud. We adopt the velocity gradient $dv/dr$ = 1.2 km s$^{-1}$ / 5.4 pc = 0.22 km s$^{-1}$ pc$^{-1}$, where $dv$ is the FWHM linewidth of $^{13}$CO($J$ = 2--1) and $dr$ is determined as an effective radius of the area whose integrated $^{13}$CO($J$ = 2--1) intensity exceeds half of the peak. We also assumed the abundance ratios of [$^{12}$CO/H$_2$] $= 8 \times 10^{-6}$ and [$^{12}$CO/$^{13}$CO] = 35, following the previous GMC studies of SMC N27 and N83C \citep{1999A&A...344..817H,2017ApJ...844...98M}. {Therefore, we} adopt $X/(dv/dr) = 4 \times 10^{-5}$ (km s$^{-1}$ pc$^{-1}$)$^{-1}$, where $X$ is the abundance ratio of [$^{12}$CO/H$_2$].

Figure \ref{fig3}(a) shows CO profiles toward the GMC LIRS~36A. Each spectrum was smoothed to match the beam size of the ALMA $^{13}$CO($J$ = 2--1) data ($\Delta\theta \sim$$30''$). We find the intensity ratios of $^{12}$CO($J$ = 3--2)/$^{12}$CO($J$ = 2--1) (hereafter $R_\mathrm{CO32/CO21}$) $\sim$0.86, $^{12}$CO($J$ = 3--2)/$^{13}$CO($J$ = 2--1) (hereafter $R_\mathrm{CO32/13CO21}$) $\sim$4.97, and $^{13}$CO($J$ = 2--1)/$^{12}$CO($J$ = 2--1) (hereafter $R_\mathrm{13CO21/CO21}$) $\sim$0.17. The results of the LVG analysis are shown in Figure \ref{fig3}(b). The red, green, and blue lines represent $R_\mathrm{CO32/CO21}$, $R_\mathrm{CO32/13CO21}$, and $R_\mathrm{13CO21/CO21}$, respectively. The errors---as shown in shaded areas for each ratio---are estimated with a $1\sigma$ noise level for each spectrum and by assuming the relative calibration error of $3\%$ for the ALMA data and $5\%$ for the ASTE data. Thanks to the low noise fluctuation and high calibration accuracy of the ALMA data, we finally obtained the number density of molecular hydrogen, $n$(H$_2) = 1500^{+600}_{-300}$ cm$^{-3}$, and the kinematic temperature, $T_\mathrm{kin} = 72^{+50}_{-37}$ K, which are roughly consistent with previous studies \citep[e.g.,][]{2007AA...471..561N}. The high $T_\mathrm{kin}$ is consistent with the heating due to the strong UV radiation of LIN~78.

\subsection{CO 3--2/1--0 Intensity Ratio}\label{ratio}
\begin{figure}[]
\begin{center}
\includegraphics[width=\linewidth]{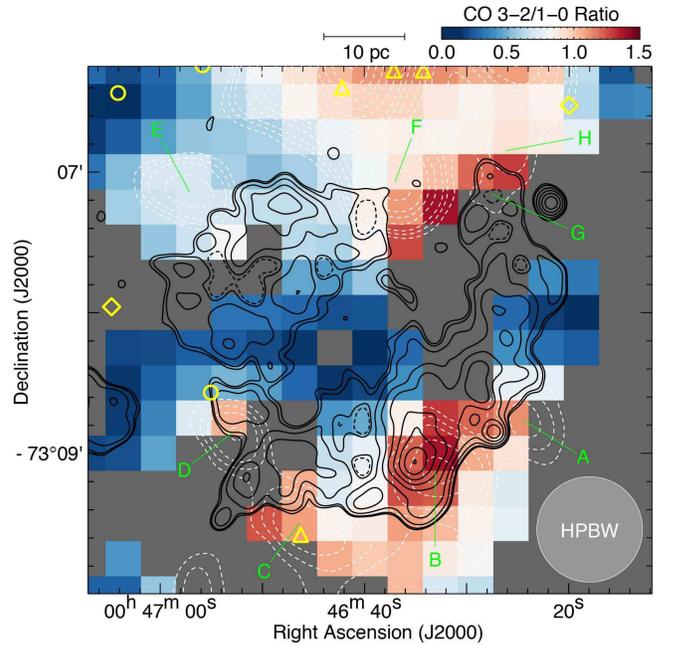}
\caption{Intensity ratio map of $^{12}$CO($J$ = 3--2) / $^{12}$CO($J$ = 1--0) using the ASTE and Mopra. Both of the data were smoothed with a Gaussian kernel to an effective beam size of $45''$. The beam size is also shown in bottom right corner of the figure. The velocity range is from $V_\mathrm{LSR}$ = 117.6 to 129.7 km s$^{-1}$. Black and white dashed contours represent the X-ray intensity and $^{12}$CO($J$ = 3--2) integrated intensity, respectively. The contour levels and symbols are the same as in Figure \ref{fig1}. The gray areas represent that the $^{12}$CO($J$ = 3--2) and/or $^{12}${CO}($J$ = 1--0) data show the low significance of $\sim$3$\sigma$ or lower. {The labels of CO clouds A--H are also shown.}}
\label{fig4}
\end{center}
\vspace*{0.5cm}
\end{figure}

We investigate the physical condition of the molecular clouds A--H by using ASTE $^{12}$CO($J$ = 3--2) and Mopra $^{12}$CO($J$ = 1--0) datasets. Figure \ref{fig4} shows the intensity ratio map of $R_\mathrm{CO32/CO10}$ toward RX~J0046.5$-$7308. Each CO data {{set}} was smoothed to match the beam size of the Mopra $^{12}$CO($J$ = 1--0) data ($\Delta\theta \sim$45$''$). We present only regions {{in which}} both {{emission}} were significantly detected (a $3\sigma$ level or higher). We find that the {high intensity} ratios of $R_\mathrm{CO32/CO10} > 1.2$ are seen toward southwest (clouds A--B), southeast (clouds C--D), and northwest (clouds F--G) of the SNR. By contrast, CO clouds E and H show a relatively low intensity ratios of $R_\mathrm{CO32/CO10} \sim$0.7 or lower.

\subsection{H{\sc i} Distribution}\label{hi_cloud}
\begin{figure*}[]
\begin{center}
\includegraphics[width=\linewidth]{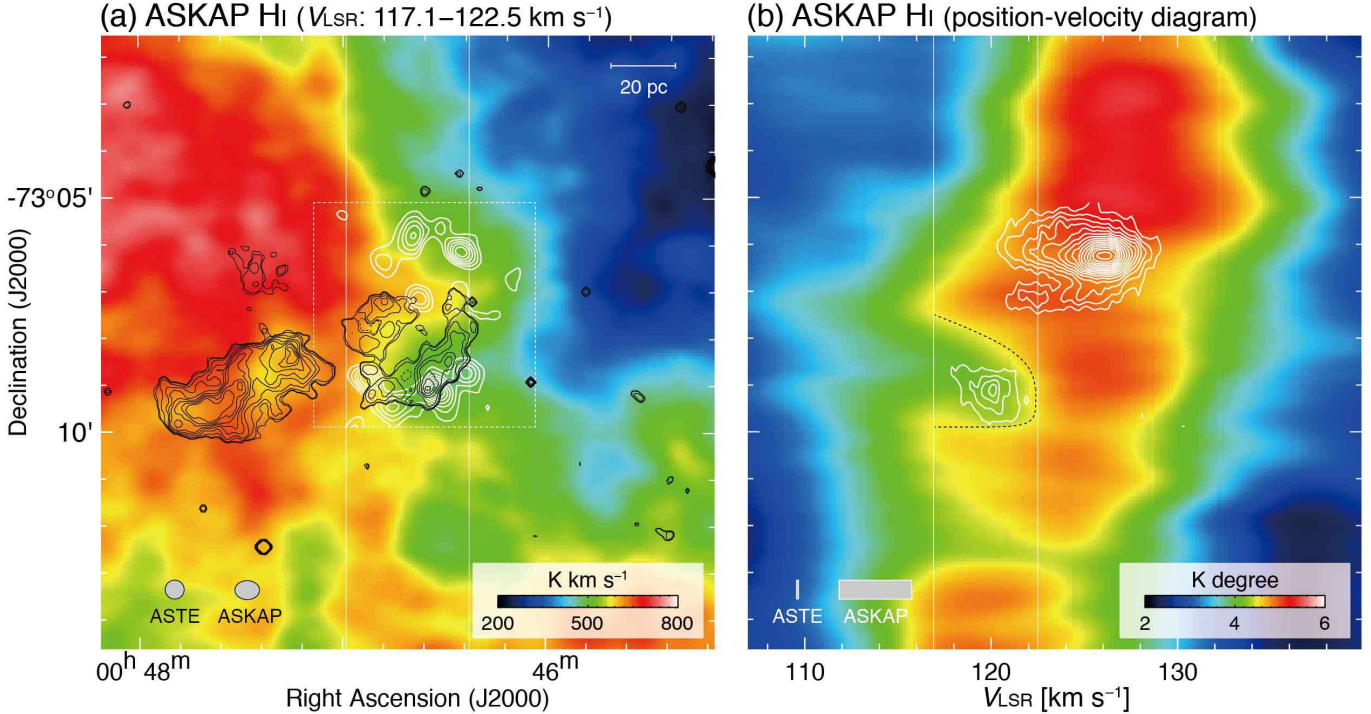}
\caption{(a) Integrated intensity map of the ASKAP H{\sc i} overlaid with the {\it Chandra} X-ray median-filtered intensity (black contours) and ASTE $^{12}$CO($J$ = 3--2) integrated intensity (white contours). The integration velocity range of CO and {H}{\sc i} is from $V_\mathrm{LSR}$ = 117.1 to 122.5 km s$^{-1}$. The contour levels of X-rays are the same as in Figure \ref{fig1}. The lowest contour and contour intervals of CO are 0.35 K km s$^{-1}$ and 0.2 K km s$^{-1}$, respectively. The scale bar and beam size are also shown in top right corner and bottom left corner, respectively. {The solid vertical lines indicate the integration range of the position-velocity diagram in Figure \ref{fig5}(b).} (b) Position--velocity diagram of H{\sc i} intensity overlaid with $^{12}$CO($J$ = 3--2) intensity contours. The integration range in R.A. is from 11\fdg59 to 11\fdg75. The lowest contour and contour intervals of CO are 0.004 K degree and 0.003 K degree, respectively. The beam size is also shown in bottom left corner. The {solid vertical lines indicate} a boundary of the H{\sc i} cavity in the position-velocity diagram (see the text).}
\label{fig5}
\end{center}
\vspace*{0.5cm}
\end{figure*}

Figure \ref{fig5}(a) shows a large-scale H{\sc i} map obtained with ASKAP superposed on the $Chandra$ X-rays and ASTE CO contours. We selected integration velocity range from 117.1 to 122.5 km s$^{-1}$, which is covered the molecular clouds A--D, F, and G showing the {high intensity} ratio of $R_\mathrm{CO32/CO10} > 1.2$ (see also Figure \ref{fig4} and Table \ref{table}). There is an H{\sc i} intensity gradient increasing from west to east, and the brightest feature has intensities above $\sim$700 K km s$^{-1}$. In the direction of the SNR, we find a cavity-like structure of H{\sc i}. The boundary of H{\sc i} cavity is nicely along the X-ray shell, especially in the northeastern half. On the other hand, the southwestern SNR shell has no prominent H{\sc i} structure where the molecular clouds A and B are located.

Figure \ref{fig5}(b) shows a position--velocity diagram of H{\sc i} and CO. The H{\sc i} clouds lie on the velocity range from $\sim$115 to $\sim$135 km s$^{-1}$. {The integration range in R. A. is from 11\fdg59 to 11\fdg75, which roughly corresponds to the diameter of the SNR.} We find that the H{\sc i} is hollowed out at the position of ($V_\mathrm{LSR}$, $\delta_\mathrm{J2000}$) $\sim$ (120 km s$^{-1}$, $-73\degr09\arcmin$). The hollowed out region is over the velocity range that shown in Figure \ref{fig5}(a) ($V_\mathrm{LSR} \sim$ 117.1--122.5 km s$^{-1}$). Moreover, the spatial extent of the hollowed out region is {roughly} similar to the shell size of the SNR. We also confirmed the observational trends using an archival H{\sc i} dataset (an angular resolution of $\sim100''$ and a velocity resolution of $\sim1.65$ km s$^{-1}$) obtained with the Australia Telescope Compact Array and the Parkes telescope published by \cite{1999MNRAS.302..417S}.

\subsection{X-Ray Spectral Analysis}\label{sec:xspec}
\begin{figure}[]
\begin{center}
\includegraphics[width=\linewidth]{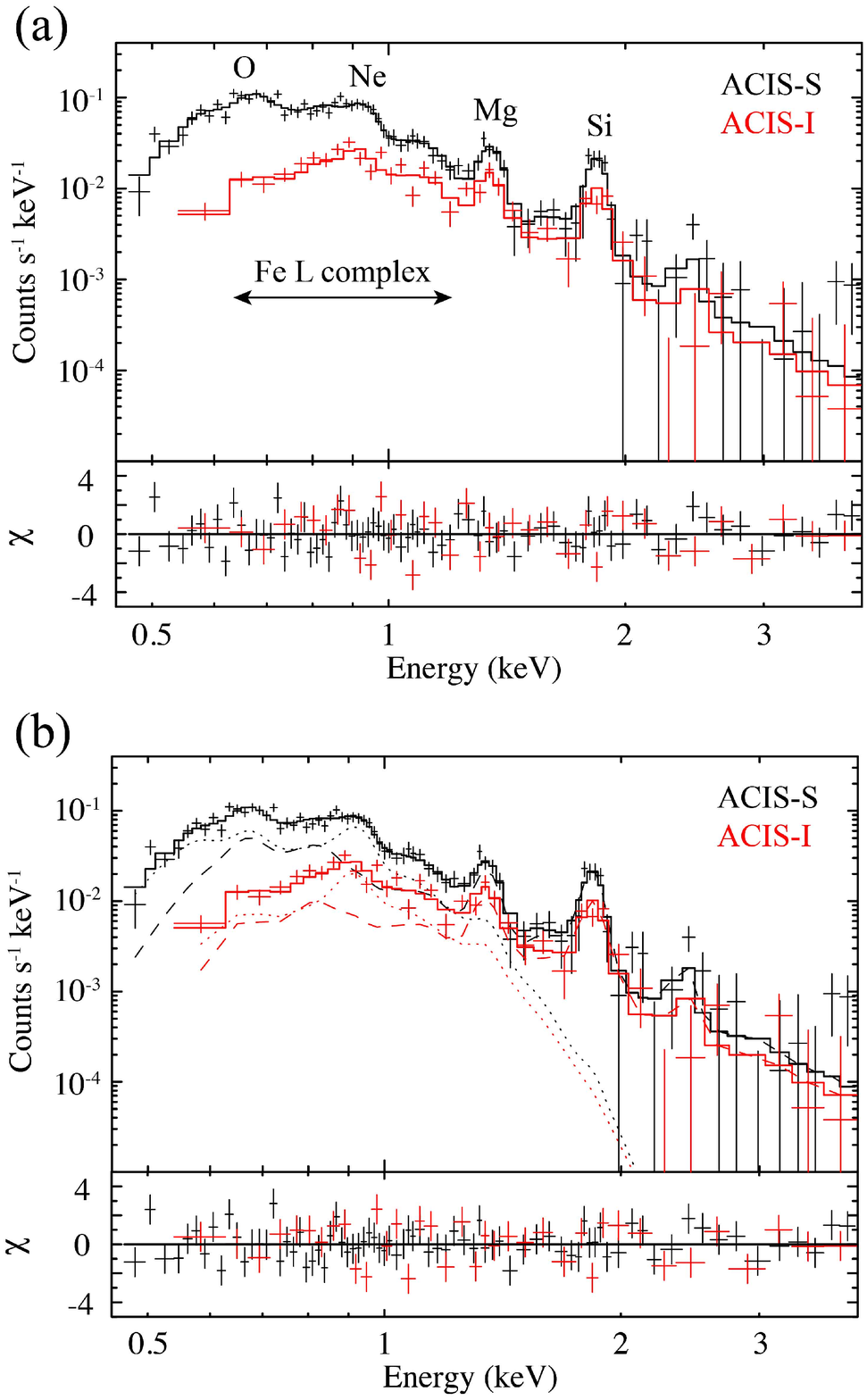}
\caption{(a) Background-subtracted {ACIS-S (black) and ACIS-I (red) spectra} of the source region (crosses in the top panel) with the best-fit NEI model (solid lines). The residuals from the best-fit model are denoted by the crosses in the bottom panel. (b) Same as (a) but the best-fit model (solid lines) is the NEI (dashed lines) + CIE (dotted lines).}
\label{fig:whole_spe}
\end{center}
\end{figure}

We extracted {ACIS-S and ACIS-I spectra} from the source region indicated in Figure~\ref{fig1}. Background is selected as a source-free region centered at ($\alpha_{\rm J2000}$, $\delta_{\rm J2000}$) $\sim$ ($00^{\rm{h}}56^{\rm{m}}59\farcs7, -73\degr 05\arcmin 04\farcs$8) with a radius of $83\farcs7$. Figure~\ref{fig:whole_spe} shows the background-subtracted {spectra} of RX~J0046.5-7308. We find O, Ne, Mg, and Si K-shell line emission and an Fe L complex, indicating that these atoms are highly ionized.


According to a previous study with the {{\it{XMM-Newton}}} \citep{2004AA...421.1031V}, the SNR spectrum can be well reproduced by an NEI plasma model with the electron temperature as $kT_e = 1.51 \pm 0.48$~keV and the ionization parameter as $n_et = (1.0 \pm 4.5) \times 10^{10}~\rm cm^{-3}$ s. Following them, we fitted the spectrum with an NEI model (VVRNEI in XSPEC). We separately set absorption column densities in the Milky Way ($N_{\rm H, MW}$) and the SMC ($N_{\rm H, SMC}$). We used the Tuebingen-Boulder ISM absorption model \citep[TBabs;][]{Wilms2000} and fixed $N_{\rm H, MW}$ at $6.0 \times 10^{20}~\rm cm^{-2}$ \citep{Dickey1990}. We allowed $kT_e$, $n_{e}t$, and the volume emission measure (VEM $= \int n_e n_p dV$) to vary. The abundances of O, Ne, Mg, Si and Fe are free parameters. {We linked the abundance of Ni and those of S, Ar, and Ca to Fe and Si, respectively.} The other elements were fixed to the SMC values in the literature {\citep[He = 0.85, C = 0.10, N = 0.09, and others = 0.20;][]{2019AJ....157...50D}}.

Figure~\ref{fig:whole_spe}(a) and Table~\ref{tab:X-ray_fit} show the fitting results and the best-fit parameters, respectively. {The spectra are} reproduced by the NEI model with {$kT_e = 1.03 _{-0.43}^{+1.03}$~keV and $n_et = (7.1 _{-2.0}^{+5.7}) \times 10^{9}~\rm cm^{-3}$ s ($\rm \chi^2/d.o.f. = 144.2/106$)}. Obtained metal abundances of O, Ne, Mg, Si, and Fe are significant higher than the SMC values {\citep[O = 0.23, Ne = 0.20, Mg = 0.21, Si = 0.30, and Fe = 0.22;][]{2019AJ....157...50D}}, suggesting that the SNR plasma possibly originates from the SN ejecta.

\begin{deluxetable}{llcc}[]
\tablecaption{Best-fit X-Ray Spectral Parameters
\label{tab:X-ray_fit}}
\tablehead{
\colhead{Component} &
\colhead{Parameter (Unit)} &
\colhead{NEI} &
\colhead{NEI+CIE} 
} 
\startdata
	Absorption & $N_{\rm H, MW}~(10^{21}~\rm cm^{-2})$ & 0.6 (fixed) & 0.6 (fixed)\\
	& $N_{\rm H, SMC}~(10^{21}~\rm cm^{-2})$ & 4.6 $_{-1.2}^{+0.9}$ & 8.4 $_{-0.8}^{+0.5}$\\
	NEI & $kT_e~\rm (keV)$ & 1.03 $_{-0.43}^{+1.03}$ & 1.09 $_{-0.38}^{+0.76}$\\
	& $Z_{\rm O}~\rm (solar)$ & 1.6 $_{-0.6}^{+0.7}$ & 4.6 $_{-1.9}^{+3.8}$\\
	& $Z_{\rm Ne}~\rm (solar)$ & 1.1 $_{-0.4}^{+0.5}$ & $<$ 1.4\\
	& $Z_{\rm Mg}~\rm (solar)$ & 1.5 $_{-0.5}^{+0.7}$ & 1.8 $_{-0.7}^{+1.3}$\\
	& $Z_{\rm Si}~\rm (solar)$ & 6.8 $_{-1.8}^{+2.3}$ & 8.3 $_{-2.6}^{+5.2}$\\
	& $Z_{\rm Fe}~\rm (solar)$ & 1.1 $_{-0.6}^{+0.9}$ & 1.0 $_{-0.5}^{+0.9}$\\
	& $n_et~\rm (10^{9}~cm^{-3}~s)$ & 7.1 $_{-2.0}^{+5.7}$ & 7.7 $_{-1.8}^{+3.6}$\\
	& VEM~($\times 10^{57}~\rm cm^{-3}$) & 4.7 $_{-2.0}^{+5.7}$ & 3.7 $_{-2.2}^{+4.3}$\\
	CIE & $kT_e~\rm (keV)$ & $\cdots$ & 0.13 $_{-0.01}^{+0.02}$\\ 
	& VEM~($\times 10^{60}~\rm cm^{-3}$) & $\cdots$ & 3.9 $_{-2.0}^{+3.2}$\\ 
      \hline
      & reduced $\chi^2$ & 1.36 & 1.32\\
      & d.o.f.	 & 106 & 104\\
\enddata
\end{deluxetable}

SNRs in the Sedov phase generally {{show significant  emission from}} swept-up ISM. To estimate the contribution of the emission from the shock-heated ISM, we {attempted} to fit the {spectra} with a two-plasma model composed {of} the ISM and ejecta. For the ejecta component, we adopted an NEI model that is the same as the previous fit. For the swept-up ISM, we first applied an NEI model whose abundances are fixed to the SMC values {\citep{2019AJ....157...50D}}, but its $n_et$ becomes larger than $10^{13}~\rm cm^{-3}$ s. Therefore, we treat the plasma as a collisional ionization equilibrium (CIE) by fixing $n_et$ at $10^{13}~\rm cm^{-3}$ s. The results and the best-fit parameters are shown in Figure~\ref{fig:whole_spe}(b) and Table~\ref{tab:X-ray_fit}. The two-components model well reproduces the spectrum with $\rm \chi^2/d.o.f. = {137.5/104}$. The fit improves the reduced $\chi^2$ from the single NEI fit, but the improvement is not statistically significant with an F-test probability of {0.084}. The {{bright-shell}} X-ray morphology suggests that the shockwave heated the swept-up ISM and that the SNR plasma contains a large amount of the ISM. This is consistent with our result that the VEM of the ISM component is much higher than that of the ejecta component. {Therefore, we} conclude that the two-components model is more suitable than the single-NEI model.

\section{Discussion}\label{sec:discussion}
\subsection{Estimation of the age and typing of the SNR}\label{metal}
In the X-ray spectral analysis, we reproduced the SNR plasma with the NEI $+$ CIE model composed of the ejecta and the swept-up ISM. We found that the ISM plasma is in the CIE state {where} $n_et$ is larger than $2 \times 10^{12}~\rm cm^{-3}~s$ \citep[][]{Masai1994}, suggesting that RX~J0046.5$-$7308 is a middle-aged SNR.

In the Sedov--Taylor phase \citep{1959sdmm.book.....S}, the dynamical age of the SNR $t_\mathrm{dyn}$ is described by 
\begin{eqnarray}
t_\mathrm{dyn} = \frac{2 R_\mathrm{sh}}{5 V_\mathrm{sh}}
\label{eqx00}
\end{eqnarray}
where $V_\mathrm{sh}$ is the shock velocity, and $R_\mathrm{sh}$ is the radius of the SNR. We here adopt $R_\mathrm{sh} \sim 22$ pc, which is {the} mean radius of the X-ray shell size \citep{2008A&A...485...63F}. Assuming the ion--electron temperature equilibration, $V_\mathrm{sh}$ can be derived as follows:
\begin{eqnarray}
V_\mathrm{sh} = \sqrt{\frac{16 k_\mathrm{B} T_\mathrm{sh}}{3 \mu m_\mathrm{H}}}
\label{eqx01}
\end{eqnarray}
where $\mu = 0.604$ is mean atomic weight, $k_\mathrm{B}$ is Boltzmann's constant, and $T_\mathrm{sh}$ is the obtained shock temperature of {$\sim0.13_{-0.01}^{+0.02}$ keV} (see Table \ref{tab:X-ray_fit}). We then obtain the shock velocity of {$V_\mathrm{sh} = 332^{+24}_{-13}$ km s$^{-1}$} and the dynamical age {$t_\mathrm{dyn} = 26000_{-2000}^{+1000}$ yr}, which are roughly consistent with the previous X-ray study \citep{2004AA...421.1031V}. Since the estimated age is long, the reverse shock probably reached the center of the remnant and heated the ejecta from the surface to the core.

\begin{figure}[]
\begin{center}
\includegraphics[width=\linewidth]{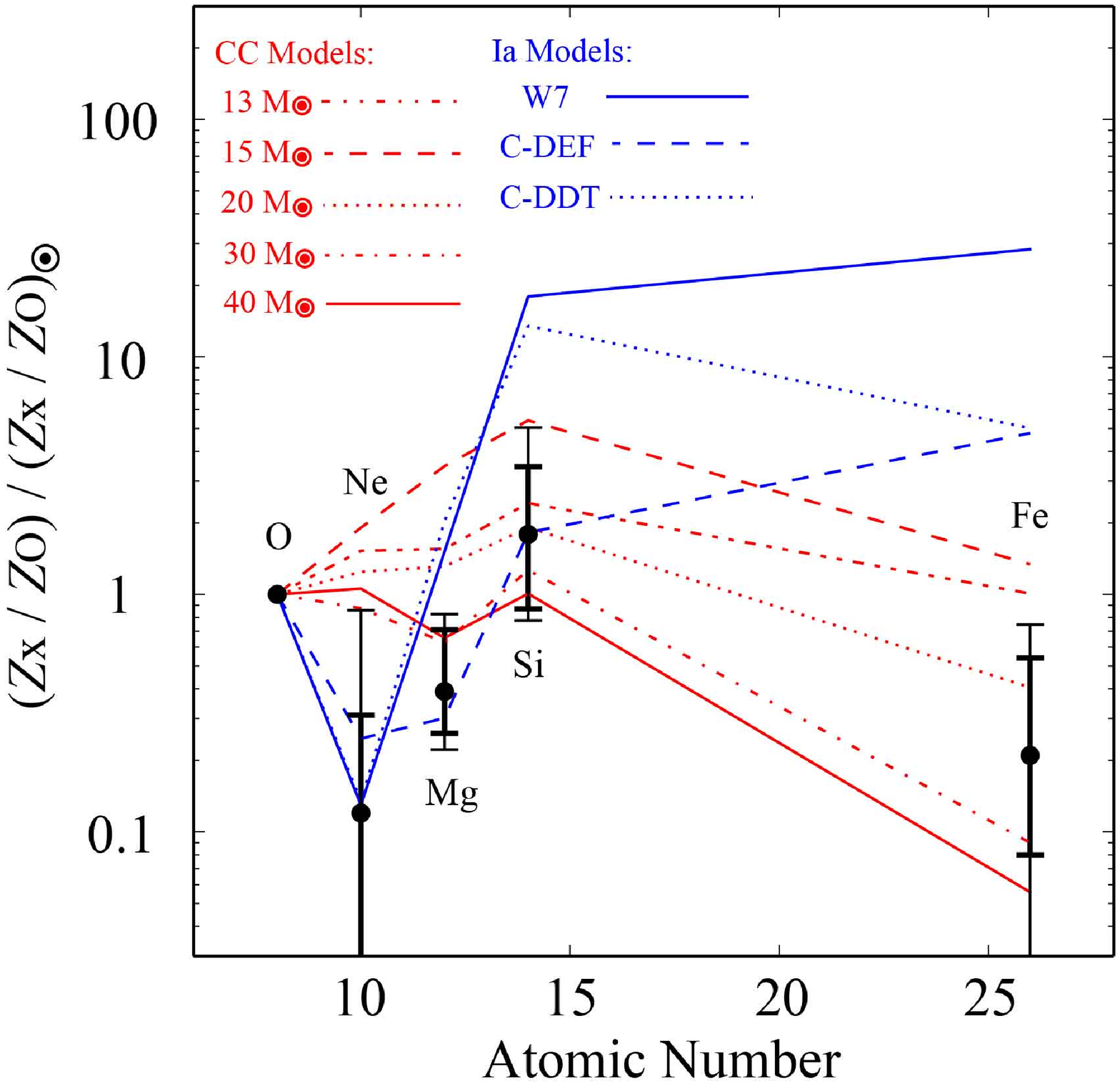}
\caption{Abundance ratios of the ejecta of Ne, Mg, Si, S, and Fe to O (circles) obtained from the NEI $+$ CIE model, relative to the abundance ratios of \cite{Wilms2000}. {The bold errors are given by the NEI $+$ CIE fit. The fine errors include $1\sigma$ confidence levels given by the models whose ISM abundances are twice or half.} The blue lines indicate the Ia SN models (W7, \citeauthor{Nomoto1984} \citeyear{Nomoto1984}, C-DEF, and C-DDT, \citeauthor{Maeda2010} \citeyear{Maeda2010}). The red lines denote the CC SN models with different progenitor masses \citep{Kobayashi2006}.}
\label{fig:SN_type}
\end{center}
\end{figure}

The spectral analysis also revealed the abundances of O, Ne, Mg, Si, and Fe of the ejecta component.  
In order to identify the SN type of RX~J0046.5$-$7308, we compared the abundance pattern of the ejecta and those derived from theoretical simulations. {Figure~\ref{fig:SN_type} shows abundance ratios of Ne, Mg, Si, and Fe to O with $1\sigma$ error confidence levels (bold errors). Abundance patterns of Ia SN models \citep{Nomoto1984,Maeda2010} and CC SN models \citep{Kobayashi2006}  are also shown in the figure.} {The ratio of Fe/O has been shown to be a better estimator of the progenitor mass than the more commonly used X/Si ratio, which is usually the only one accessible for the heavily absorbed Galactic SNRs \citep[][]{2018ApJ...863..127K}.} {Although it is difficult to estimate the SN type from the ratios of Mg/O and Si/O, those of Ne/O and Fe/O show contradictory results. The ratio of Fe/O clearly indicates that the remnant is {{that of}} a CC SN with a heavier progenitor mass of $\gtrsim 20~M_{\sun}$. The ratio of Ne/O, on the other hand, suggests an Ia origin but the Ne abundance of the ejecta has large uncertainties for the fluctuation of metal abundances of the ISM component.}

{To investigate the effect of the fluctuation, we additionally analyzed the X-ray spectra with the same NEI + CIE models as the previous fit but their ISM abundances are fixed to twice as or half of the SMC values \citep{2019AJ....157...50D}. Both of the models can reproduce the spectra with $\chi^2/d.o.f.~ (Z_{\rm SMC, twice}) = 138.0/104$ and $\chi^2/d.o.f.~ (Z_{\rm SMC, half}) = 138.5/104$, and we obtained the abundance ratios of the ejecta component including the fluctuation of the ISM abundances (see the fine error bars in figure~\ref{fig:SN_type}). Although the ratio of Fe/O remains the range of the CC models, that of Ne/O was allowed a CC origin with a progenitor mass of $\gtrsim 30~M_{\sun}$. Since all the ratios allowed the CC model with the mass of $\gtrsim 30~M_{\sun}$, we conclude that RX~J0046.5$-$7308 is {{of}} a CC SN origin with {{a high}} progenitor mass. The {{higher}} mass is consistent with the stellar age interpretation of \citet{2019ApJ...871...64A}, based on local stellar population alone.}

\subsection{Molecular and Atomic clouds associated with the SMC SNR~RX~J0046.5$-$7308}\label{section4.2}
Over the last three decades, we learned how to identify shocked gas clouds associated with SNRs except for morphological aspects. For the middle-aged SNRs ($\sim$10,000 yr old), there are two pieces of evidence for the shocked molecular cloud. One is the {high intensity ratio} between the $^{12}$CO($J$ = 3--2) and $^{12}$CO($J$ = 1--0) transitions. This ratio is a good indicator of the degree of the rotational excitation in CO because the $J$ = 3 state lies at 33.2 K from the $J$ = 0 ground state, which is $\sim$28 K above the $J$ = 1 state at 5.5 K. \cite{1999PASJ...51L...7A} demonstrated that a shocked molecular cloud in W28 has a {high intensity ratio} of $\sim$1.2--2.8 with OH masers (1720.5~MHz), whereas an unshocked cloud shows a low intensity ratio of $\sim$0.4--0.7. Similar trends were found in both the Galactic and Magellanic SNRs (e.g., IC~443, \citeauthor{1994A&A...283L..25W} \citeyear{1994A&A...283L..25W}; Kesteven~79, \citeauthor{2018ApJ...864..161K} \citeyear{2018ApJ...864..161K}; LMC SNR N49, \citeauthor{2018ApJ...863...55Y} \citeyear{2018ApJ...863...55Y}). The second is a broad-line profile of CO emission. A shocked molecular cloud can be accelerated about a few 10 km s$^{-1}$ if the shock-interacting time is long enough. The accelerated clouds are, therefore, observed as broad-line profiles in the CO emission \citep[e.g.,][]{1977ApJ...216..440W,1998ApJ...505..286S,2013ApJ...768..179Y}.

Recently, \cite{2017JHEAp..15....1S} {presented a hole-like (or hollowed out) structure of H{\sc i} in the position-velocity diagram toward the young SNR RCW~86 ($\sim$1800 yr old) consistent with what has also been observed in Galactic SNRs \citep[e.g.,][]{1990ApJ...364..178K,1991ApJ...382..204K}.} The hole-like structure means an expanding gas motion, also called the ``wind-blown shell,'' created by gas winds from the progenitor system: e.g., stellar winds from a massive progenitor or accretion winds (also referred to as ``disk wind'') from a single-degenerated progenitor system of the Type Ia explosion. The size of the wind-blown shell generally coincides with the diameter of the SNR because the free-expansion phase is short enough. Subsequent studies confirmed this idea in both the Galactic and Magellanic SNRs (e.g., Kesteven~79, \citeauthor{2018ApJ...864..161K} \citeyear{2018ApJ...864..161K}; LMC SNR N103B, \citeauthor{2018ApJ...867....7S} \citeyear{2018ApJ...867....7S}; \citeauthor{arXiv:1903.03226} \citeyear{arXiv:1903.03226}).

For RX J0046.5$-$7308, we first claim that the molecular clouds A, B, F, and G are most likely interacting with the shockwaves. The physical relations between the molecular clouds and shockwaves are supported by the {high intensity ratios} of $R_\mathrm{CO32/CO10} > 1.2$ without {an external stellar heating source, such as} infrared sources and/or massive stars, indicating that the shock heating occurred. {The value of $R_\mathrm{CO32/CO10} > 1.2$ is also consistent with the previous studies of shock-heated molecular clouds associated with the Galactic / Magellanic SNRs} \citep[e.g.,][]{1994A&A...283L..25W,1999PASJ...51L...7A,2018ApJ...864..161K,2018ApJ...863...55Y}. In the morphological aspects, these molecular clouds are {{located}} nicely along the X-ray shell (see Figure \ref{fig1}(b)). The southwestern shell is slightly deformed along the CO clouds A and B with the brightest X-ray peak, indicating that the shock ionization occurred \citep[e.g.,][]{2017JHEAp..15....1S}. On the other hand, we could not find the broad-line profiles of CO emission in the shocked molecular clouds. This is inconsistent with the old dynamical age of $\sim$26000 yr. It is possible that the sensitivity and angular resolution of ASTE CO data are not high enough to detect the broad-line profiles of CO emission. Further ALMA observations with high-angular resolution ($\sim 0.1$ pc) and high sensitivity are needed to detect the shock-accelerated molecular clouds in RX J0046.5$-$7308.

Next, we argue that H{\sc i} clouds at $V_\mathrm{LSR} \sim$117.1--122.5 km s$^{-1}$ and the molecular clouds C and D are also associated with the SNR in addition to clouds A, B, F, and G. The hollowed out structure in the position-velocity diagram of H{\sc i} is likely an expanding gas motion originated from stellar winds from a massive progenitor. The expanding velocity $\Delta V$ is estimated to be $\sim$3 km s$^{-1}$, which is roughly consistent with the Galactic CC SNR Kesteven~79 \citep[$\Delta V$$\sim$4 km s$^{-1}$;][]{2018ApJ...864..161K}. In fact, the metal abundances of the SNR favor the CC explosion (see Section \ref{metal}), which can create {{a}} wind-blown bubble. If the interpretation is correct, the radial velocity of shock-interacting gas is to be $\sim$117.1--122.5 km s$^{-1}$, corresponding to the velocity range of the expanding H{\sc i} shell. The molecular clouds C and D are, therefore, associated with the SNR because their radial velocites are $\sim$120 km s$^{-1}$. In light of these considerations, we conclude that the molecular clouds A--D, F, and G and H{\sc i} {gas} at the velocity range of $\sim$117.1--122.5 km s$^{-1}$ are associated with the SNR RX~J0046.5$-$7308.

\subsection{Prospects for $\gamma$-ray observations}
We also discuss future prospects for $\gamma$-ray observations toward RX~J0046.5$-$7308. $\gamma$-rays from middle-aged SNRs are mainly produced by two mechanisms: hadronic and leptonic processes. For the hadronic process, the interaction between cosmic-ray and interstellar protons creates a neutral pion that quickly decays into two $\gamma$-ray photons. Therefore, it is also referred to as the pion-decay $\gamma$-rays. For the leptonic process, cosmic-ray electron energizes a low-energy photon (e.g., cosmic-microwave background and infrared photons) into the $\gamma$-ray energy through the inverse Compton scattering. The cosmic-ray electrons also emit $\gamma$-rays via nonthermal Bremsstrahlung. There are two ways to distinguish between the hadronic and leptonic processes. One is searching for the spectral-break (or refer to as ``pion-decay bump'') of hadronic $\gamma$-rays below $\sim$200~MeV \citep[e.g.,][]{2011ApJ...742L..30G,2013Sci...339..807A}. Since each neutral pion having an energy of 67.5 MeV in the rest frame, the hadronic $\gamma$-ray number spectrum shows symmetry about 67.5 MeV in a log-log representation \citep{1971NASSP.249.....S}. The hadronic $\gamma$-ray spectrum, $F(\varepsilon)$, therefore, rises steeply below $\sim$200~MeV in the $\varepsilon^2 F(\varepsilon)$ representation. The other is probing the good spatial correspondence between the $\gamma$-rays and interstellar protons, which is an essential signature of the hadronic $\gamma$-rays \citep[][]{2003PASJ...55L..61F,2012ApJ...746...82F,2017ApJ...850...71F,2013ASSP...34..249F,1994A&A...285..645A,2008A&A...481..401A,2012PASJ...64....8H,2013ApJ...768..179Y,2012MNRAS.422.2230M,2013MNRAS.434.2188M,2013PASA...30...55M,2018MNRAS.474..662M,2018ApJ...866...76M,2018MNRAS.480..134M,2014ApJ...788...94F,2017MNRAS.464.3757L,2019MNRAS.483.3659L,2017MNRAS.468.2093D,2017ApJ...843...61S,2018arXiv180510647S,2018ApJ...864..161K}. For RX~J0046.5$-$7308, the eight molecular clouds have the potential to be detected by  TeV $\gamma$-rays using the Cherenkov Telescope Array with deep exposure. The $\gamma$-rays produced by escaped cosmic-ray protons may be detected from the nearby GMC because the physical conditions of the GMC LIRS 36A---size, mass, and separation from the SNR---are similar to the Galactic $\gamma$-ray SNR W28 \citep[e.g.,][]{2008A&A...481..401A,2010A&A...516L..11G,2010ApJ...718..348A}. RX~J0046.5$-$7308 and its surroundings possibly provide us with the best laboratory to search for the hadronic $\gamma$-rays originated from the shock-accelerated and/or escaped cosmic-ray protons in the SMC.

\section{Conclusions}\label{sec:conclusions}
We have presented new $^{12}$CO($J$ = 1--0, 2--1, 3--2) and $^{13}$CO($J$ = 2--1) observations and H{\sc i} toward the SMC SNR RX~J0046.5$-$7308 using the ASTE, Mopra, ALMA, and ASKAP. The primary conclusions are summarized as below.

\begin{enumerate}
\item We {{discovered}} eight molecular clouds, A--H, along the X-ray shell of the SNR, which are significantly detected by $^{12}$CO($J$ = 3--2) line emission obtained with the ASTE. The typical cloud size and mass are $\sim$10--15 pc and $\sim$1000--3000 $M_\sun$, respectively. The X-ray shell is slightly deformed and has the brightest peak in the southwestern shell where the molecular clouds A and B are associated. The four molecular clouds A, B, F, and G show {high intensity ratios} of $R_\mathrm{CO32/CO10} > 1.2$ without  infrared sources and/or massive stars. These results provide the first evidence for the shock-heated molecular clouds in the SMC.
\item {{A cavity-like structure in H{\sc i}}} is found toward the SNR, which is also observed as a hollowed out structure in the position-velocity diagram. The hollowed out structure of H{\sc i} is likely an expanding gas motion with an expanding velocity of $\sim$3 km s$^{-1}$, which {{was created}} by stellar winds {{of the}} massive progenitor. If the interpretation is correct, the radial velocity of shock-interacting gas is to be $\sim$117.1--122.5 km s$^{-1}$, including the peak radial velocities of molecular clouds C and D. We finally conclude that the molecular clouds A--D, F, and G and H{\sc i} {gas} within a velocity range of 117.1--122.5 km s$^{-1}$ are to be associated with the SNR RX~J0046.5$-$7308.
\item The X-ray spectral analysis revealed that the SNR plasma can be reproduced {{by an}} NEI $+$ CIE model composed {of} the ejecta and the swept-up ISM {{emission}}. Assuming the Sedov--Taylor phase, the dynamical age of the SNR is estimated to be {$26000_{-2000}^{+1000}$ yr}, which is roughly consistent with the previous X-ray studies \citep{2004AA...421.1031V}. We also obtained the abundances of O, Ne, Mg, Si, and Fe of the ejecta. {The ratios of the metal abundances to O suggest that the SNR originated from a CC SN with a heavy progenitor mass of $\gtrsim 30~M_{\sun}$}.
\item To derive the physical conditions of the GMC LIRS~36A, we carried out the LVG analysis using the $^{12}$CO($J$ = 2--1, 3--2) and $^{13}$CO($J$ = 2--1) datasets obtained with the ASTE and ALMA. We obtained the number density of molecular hydrogen of $1500^{+600}_{-300}$ cm$^{-3}$ and the kinematic temperature of $72^{+50}_{-37}$ K. Since the GMC is located far from the shell boundary of the SNR, the GMC may not be affected by the SNR shockwaves. The high kinematic temperature is, therefore, due to the heating by the massive exciting star LIN~78 in the H{\sc ii} region DEM~S23.
\item We found that the physical conditions of the molecular clouds toward RX~J0046.5$-$7308 are similar to that of the Galactic $\gamma$-ray SNR W28. We, therefore, suggest that RX~J0046.5$-$7308 and its surrounding molecular clouds provide us with the best laboratory to search the pion-decay $\gamma$-rays that originated from the shock-accelerated and/or escaped-cosmic ray protons in the SMC. 
\end{enumerate}

\acknowledgments
{{\small This paper makes use of the following ALMA data: ADS/JAO.ALMA \#2015.1.00196.S. ALMA is a partnership of ESO (representing its member states), NSF (USA), and NINS (Japan), together with NRC (Canada) and NSC and ASIAA (Taiwan) and KASI (Republic of Korea), in cooperation with the Republic of Chile. The Joint ALMA Observatory is operated by ESO, AUI/NRAO and NAOJ. The ASTE radio telescope is operated by NAOJ. The Mopra radio telescope and Australian SKA Pathfinder (ASKAP) are part of the Australia Telescope National Facility which is managed by CSIRO. The University of New South Wales Mopra Spectrometer Digital Filter Bank used for these Mopra observations was provided with support from the Australian Research Council, together with the University of New South Wales, the University of Adelaide, University of Sydney, Monash University, and the CSIRO. Operation of the ASKAP is funded by the Australian Government with support from the National Collaborative Research Infrastructure Strategy. The ASKAP uses the resources of the Pawsey Supercomputing Centre. Establishment of the ASKAP, the Murchison Radio-astronomy Observatory, and the Pawsey Supercomputing Centre are initiatives of the Australian Government, with support from the Government of Western Australia and the Science and Industry Endowment Fund. We acknowledge the Wajarri Yamatji people as the traditional owners of the Observatory site. The scientific results reported in this article are based on data obtained from the {\it Chandra} Data Archive (Obs ID: 3904, 14674, 15507, and 16367). This research has made use of software provided by the {\it Chandra} X-ray Center (CXC) in the application packages CIAO (v.4.10). This study was financially supported by Grants-in-Aid for Scientific Research (KAKENHI) of the Japanese Society for the Promotion of Science (JSPS, grant Nos. 16K17664, 18J01417, and 19K14758). H.S. was supported by ``Building of Consortia for the Development of Human Resources in Science and Technology'' of Ministry of Education, Culture, Sports, Science and Technology (MEXT, grant No. 01-M1-0305). H.M. was supported by World Premier International Research Center Initiative (WPI). K. Tokuda was supported by NAOJ ALMA Scientific Research (grant No. 2016-03B). H.S. was also supported by the ALMA Japan Research Grant of NAOJ Chile Observatory (grant Nos. NAOJ-ALMA-201 and NAOJ-ALMA-208).}}
\software{CASA \citep[v 4.5.3.:][]{2007ASPC..376..127M}, CIAO \citep[v 4.10:][]{2006SPIE.6270E..1VF}.}

\end{document}